\documentclass{article}
\usepackage{color}
\usepackage{graphicx}
\usepackage{amsmath, amsthm}
\usepackage{mathtools}
\usepackage{natbib}
\usepackage{url}
\usepackage{soulutf8}
\RequirePackage[colorlinks,citecolor=blue,urlcolor=blue]{hyperref}
\usepackage{xcolor}
\usepackage{chngcntr}
\usepackage{subfig}
\usepackage{hyperref}
\usepackage{algorithm}
\usepackage{algpseudocode}
\usepackage{setspace}
\newtheorem{theorem}{Theorem}
\newtheorem{corollary}{Corollary}
\usepackage{xfrac}
\definecolor{forest}{rgb}{0.133,0.545,0.133}
\usepackage{multirow}
\usepackage{amsfonts}

\usepackage{enumitem}
\usepackage{amssymb}

\evensidemargin=\oddsidemargin
\usepackage{etoolbox}

\newif\ifabbreviation
\pretocmd{\thebibliography}{\abbreviationfalse}{}{}
\AtBeginDocument{\abbreviationtrue}

\begin{document}
	\newcommand{\bb}{\boldsymbol{\beta}}

	\title{Design of Bayesian A/B Tests Controlling False Discovery Rates and Power}


	\author{Luke Hagar\footnote{Luke Hagar is the corresponding author and may be contacted at \url{lmhagar@uwaterloo.ca}.} \hspace{35pt} Nathaniel T. Stevens$^{\dagger}$ \bigskip \\ 
 $^*$\textit{Centre for Clinical Research, The University of Queensland} \\ $^{\dagger}$\textit{Department of Statistics \& Actuarial Science, University of Waterloo}}

	\date{}

	\maketitle

	\begin{abstract}

Businesses frequently run online controlled experiments (i.e., A/B tests) to learn about the effect of an intervention on multiple business metrics.  To account for multiple hypothesis testing, multiple metrics are commonly aggregated into a single composite measure, losing valuable information, or strict family-wise error rate adjustments are imposed, leading to reduced power. In this paper, we propose an economical framework to design Bayesian A/B tests while controlling both power and the false discovery rate (FDR). Selecting optimal decision thresholds to control power and the FDR typically relies on intensive simulation at each sample size considered. Our framework efficiently recommends optimal sample sizes and decision thresholds for Bayesian A/B tests that satisfy criteria for the FDR and average power. Our approach is efficient because we leverage new theoretical results to obtain these recommendations using simulations conducted at only two sample sizes. Our methodology is illustrated using an example based on a real A/B test involving several metrics.

		\bigskip

	\end{abstract}

	\maketitle

	\baselineskip=19.5pt


 \section{Introduction}\label{sec:intro}

  In many contexts involving business and the social sciences, statisticians aim to assess the impact of an intervention on multiple outcomes. One prevalent context is that of online controlled experiments run by technology-based companies to improve their products and services \citep{thomke2020experimentation, luca2021power, bajpai2025extensible}. These experiments are often called A/B tests, which is an apt moniker for comparisons involving two versions of an online experience: a control version (A) and a treatment version (B). A/B tests typically monitor many metrics that summarize different aspects of how users interact with an online experience. A limited set of these metrics defines the overall evaluation criterion (OEC) used for decision making \citep{kohavi2020trustworthy}. Combining this set of metrics into one composite OEC simplifies decision making \citep{roy2001design}, but adequately summarizing the business goals for an A/B test with a single metric can present challenges. Many companies instead examine a multi-metric OEC comprised of several key scalar metrics, and so the multiple comparisons problem \citep{hochberg2009multiple} must be accounted for during A/B test design. For further background on A/B testing methods, see \citet{larsen2024statistical}.

    For each metric $k = 1, \dots, K$, A/B tests compare complementary hypotheses $H_{0, k}$ and $H_{1,k}$. A discovery is made with respect to the $k^{\text{th}}$ metric when the data support $H_{1,k}$. A/B tests with multiple metrics are often designed to control the family-wise error rate (FWER). However, strict control of the FWER often results in a loss of power. Since insufficient power in A/B tests is a common problem \citep{deng2013improving,larsen2024statistical, deng2024metric}, any design choices that further decrease power should be avoided. Thus, we control the false discovery rate (FDR): the expected proportion of discoveries that are false across a group of hypothesis tests. Whereas the FWER conservatively bounds the probability of making even one false discovery (i.e., incorrectly concluding that one $H_{1,k}$ is true), the FDR balances the prevalence of false and true positives across multiple comparisons. The development of design methods for A/B tests with FDR control is therefore crucial to enhance the average power of these tests. Average power is the expected proportion of the true $\{H_{1,k}\}_{k = 1}^K$ that are correctly supported by the data. 

The most popular methods used to control the FDR are based on frequentist $p$-values \citep{benjamini1995controlling, benjamini2001control,storey2002direct}. Given the rising prominence of Bayesian approaches to A/B testing (see e.g., \citet{deng2015objective,stevens2022comparative, deng2024metric}), this paper focuses on testing procedures facilitated via posterior probabilities. For the $k^{\text{th}}$ metric, the observed data provide evidence to support $H_{1, k}$ if the posterior probability $Pr(H_{1, k}~|~ data)$ is greater than or equal to a decision threshold $\gamma_k \in [0.5, 1)$. Given some data generation process, a sample size $n$ could be selected to ensure the average power to observe $Pr(H_{1, k}~|~ data) \ge \gamma_k$ across all true hypotheses in $\{H_{1,k}\}_{k=1}^K$ is at least $1-\beta$. The decision thresholds $\boldsymbol{\gamma} = \{\gamma_k\}_{k=1}^K$ are chosen to bound the FDR by $q \in (0,1)$. The Bayesian FDR control provided by this framework with posterior probabilities is defined and compared with the FDR control prompted by popular frequentist methods in Section \ref{sec:prelim}.

To support the design of A/B tests with non-simplistic statistical models, sample sizes and optimal decision thresholds that control the operating characteristics of average power and the Bayesian FDR can be found using intensive simulation \citep{wang2002simulation}. In general, one must simulate many samples of size $n$ according to some assumed data generation process to estimate the joint sampling distribution of $\{Pr(H_{1, k}~|~ data)\}_{k=1}^K$. This estimate of the sampling distribution of posterior probabilities determines whether the investigated sample size and decision thresholds satisfy criteria for the operating characteristics. This computationally intensive process is repeated until a suitable $(n, \boldsymbol{\gamma})$ combination is found. An economical framework to determine optimal sample sizes and decision thresholds that satisfy criteria for the Bayesian FDR and average power would expedite the design of Bayesian A/B tests with multiple metrics, making them more accessible to practitioners.

For a single test, \citet{hagar2025economical} recently put forward an economical method to assess operating characteristics throughout the sample size space using estimates of the sampling distribution of posterior probabilities at only two sample sizes. In this work, we build upon their approach to more efficiently design A/B tests involving a host of metrics controlling the Bayesian FDR and average power.  To assess the operating characteristics of multiple tests, we must account for the dependence structure in the \emph{joint} sampling distribution of posterior probabilities across hypotheses. The method from \citet{hagar2025economical} was predicated on a theoretical proxy to the (univariate) sampling distribution of posterior probabilities for a single hypothesis. As such, new theory must be developed to accommodate arbitrary dependence structures in the joint sampling distribution. 

The remainder of this article is structured as follows.  We introduce our framework for inference and the criteria under which the Bayesian FDR and average power are controlled in Section \ref{sec:prelim}.  In Section \ref{sec:proxy}, we prove novel theoretical results about a proxy to the joint sampling distribution of posterior probabilities. We develop a method in Section \ref{sec:power} that adapts these results to determine which $(n, \boldsymbol{\gamma})$ combination minimizes the sample size $n$ while satisfying criteria for average power and the Bayesian FDR. This method efficiently considers a range of sample sizes using estimates of the sampling distributions of posterior probabilities at only two values of $n$. In Section \ref{sec:ex}, we illustrate the use of our methodology with an example from a real A/B test with multiple binary metrics that have a complicated dependence structure. We conclude with a summary and discuss extensions to this work in Section \ref{sec:disc}. 

\section{Preliminaries}\label{sec:prelim}

\subsection{Problem Set-Up}

Our design framework represents data from a future sample of size $n$ as $\boldsymbol{Y}^{_{(n)}} = \{\boldsymbol{Y}_{i} \}_{i = 1}^n$. We presume there is one outcome for each of the $K$ metrics, so each observation $\boldsymbol{Y}_{i}$ has dimension $K$. The observed data are denoted by $\boldsymbol{y}^{_{(n)}}$.  For A/B tests, $\boldsymbol{Y}^{_{(n)}}$ consists of $n_j$ observations from group $j = A, B$ such that $n = n_A + n_B$. We consider fixed treatment allocation where $n_A = \lfloor cn_B \rceil$ for some $c > 0$, but this constant $c$ is not incorporated into $\boldsymbol{Y}^{_{(n)}}$. We assume that each observation in $\boldsymbol{Y}^{_{(n)}}$ is generated independently via the model $f(\boldsymbol{y}; \boldsymbol{\eta}_A)$ or $f(\boldsymbol{y}; \boldsymbol{\eta}_B)$, depending on its group assignment. We note that $\boldsymbol{\eta}_j$ represents a vector of parameters for group $j = A, B$, and we collectively denote the parameters for both groups by $\boldsymbol{\eta} = (\boldsymbol{\eta}_A, \boldsymbol{\eta}_B)$. Since \citet{hagar2025economical} considered designs with additional covariates $\{\boldsymbol{X}_{i} \}_{i = 1}^n$, our extensions to their method could also be used to design A/B tests that involve treatment effect heterogeneity \citep{larsen2024statistical}.

The targets of inference $\boldsymbol{\theta}$ for the host of metrics are specified as a function $g(\cdot)$ of the model parameters: $\boldsymbol{\theta} = g(\boldsymbol{\eta})$. The components in $\boldsymbol{\theta} = \{\theta_k\}_{k=1}^K$ can be expressed as $\theta_k = g_k(\boldsymbol{\eta})$. For the $k^{\text{th}}$ metric, we consider interval hypotheses of the form $H_{1, k}: \theta_k \in (\delta_{L, k}, \delta_{U, k})$, where $-\infty \le \delta_{L, k} < \delta_{U, k} \le \infty$. The corresponding complementary hypothesis is $H_{0, k}: \theta_k \notin (\delta_{L, k}, \delta_{U, k})$. This general notation for the interval $(\delta_{L, k}, \delta_{U, k})$ accommodates a broad suite of hypothesis tests based on superiority and practical equivalence \citep{stevens2022comparative}. We jointly refer to the interval endpoints across all hypotheses as $\boldsymbol{\delta}_L = \{\delta_{L,k} \}_{k=1}^K$ and $\boldsymbol{\delta}_U = \{\delta_{U,k} \}_{k=1}^K$.

We specify a probability model $\Psi$ that characterizes which $\boldsymbol{\eta}$ values define the data generation process. The model $\Psi$ defines scenarios with effect sizes that align with the experiment's objectives. It differs from the prior $p(\boldsymbol{\eta})$ that defines the posterior of $\boldsymbol{\theta}$ and characterizes all available pre-experimental knowledge regardless of whether it aligns with the goals of the A/B tests.  As illustrated in Section \ref{sec:ex}, defining a model $\Psi$ that is distinct from $p(\boldsymbol{\eta})$ allows practitioners to incorporate uncertainty about \emph{how many} and \emph{which} hypotheses are false in a straightforward manner.

To  compare  and  contrast  how  the  FDR  is  controlled  in  our  design  framework  and under alternative approaches, we let $V$ and $S$ be random variables that denote the number of false and true discoveries, respectively. Our design method controls the FDR such that
\begin{equation}\label{eq:fdr.1}
\mathbb{E}_{\boldsymbol{\eta} \sim \Psi}\left\{\mathbb{E}\left[\dfrac{V}{V+S}\right]\right\} \le q,
\end{equation} 
where $0 \div 0$ is defined to be 0. In this paper, we refer to the FDR on the left side of the inequality in (\ref{eq:fdr.1}) as the \emph{Bayesian} FDR. This Bayesian FDR is controlled in expectation under the assumption that $\boldsymbol{\eta} \sim \Psi$, and we discuss the implications when this assumption does not hold true in Section \ref{sec:ex}. Our Bayesian  FDR control is also defined with respect to the model $f(\boldsymbol{y};\boldsymbol{\eta})$ for the data that practitioners must select. 

The most popular frequentist method to control the FDR was proposed by \citet{benjamini1995controlling}. Their method controls the FDR such that
\begin{equation}\label{eq:fdr.2}
\inf_{\boldsymbol{\eta}}\lim_{n \rightarrow \infty}\mathbb{E}\left[\dfrac{V}{V+S}\right] \le q,
\end{equation} 
where $0 \div 0$ is again defined to be 0. The criterion in (\ref{eq:fdr.2}) holds for an arbitrary value for the parameter $\boldsymbol{\eta}$ that indexes the data generation process. Standard methods for FDR control assume that $p$-values are uniformly distributed under each $H_{0, k}$. This uniformity is often justified asymptotically using the central limit theorem, so standard methods for FDR control do not require practitioners to choose a model $f(\boldsymbol{y}; \boldsymbol{\eta})$ for the data. However, the FDR control in \citet{benjamini1995controlling} is only attained for certain dependence structures, including settings where all the hypotheses are independent. \citet{benjamini2001control} developed a method that attains the criterion in (\ref{eq:fdr.2}) for an arbitrary dependence structure, but simulation-based procedures are recommended for optimal control of the FDR for a specific dependence structure \citep{yekutieli1999resampling}. Ultimately, our design framework with Bayesian FDR control  requires users to make more choices about the data generation process and model for the data; the trade-off of using our methods is that the recommended sample sizes and decision thresholds are optimized when these choices are reasonable.

To define our notion of average power, we let $T$ be a random variable that denotes the number of true hypotheses. Our design method controls average power across all true hypotheses in $\{H_{1,k}\}_{k = 1}^K$ such that 
\begin{equation}\label{eq:avg.pwr}
\mathbb{E}_{\boldsymbol{\eta} \sim \Psi}\left\{\mathbb{E}\left[\dfrac{S}{T}\right]\right\} \ge 1 - \beta,
\end{equation} 
where $0 \div 0$ is defined to be 0. Standard design methods also control power with respect to some data generation process $\Psi$. The goal of our proposed method is to recommend a suitable $(n, \boldsymbol{\gamma})$ combination such that the criteria for the Bayesian FDR in (\ref{eq:fdr.1}) and average power in (\ref{eq:avg.pwr}) are approximately attained. We formally describe how our method approximately attains these criteria in Section \ref{sec:power}. In the next subsection, we explain why standard methods to find suitable $(n, \boldsymbol{\gamma})$ combinations are very computationally intensive and how we improve upon standard design procedures.

\subsection{A Computational Bottleneck}

Algorithm \ref{alg.int} details a standard, simulation-driven procedure to classify hypotheses as true or false for a particular $(n, \boldsymbol{\gamma})$ combination and set of design inputs. These classifications based on posterior probabilities are used to estimate the Bayesian  FDR and average power. Before explaining why design methods that repeatedly implement Algorithm \ref{alg.int} are computationally intensive, we elaborate on the choice of the model $\Psi$ that is input into this algorithm.  Note that the Bayesian  FDR is trivially 0 if all hypotheses $\{H_{1,k}\}_{k = 1}^K$ are true, and average power is trivially 0 when all $\{H_{1,k}\}_{k = 1}^K$ are false. In this paper, we therefore focus on data generation processes where each $\boldsymbol{\eta}_r \sim \Psi$ is such that some -- but not all -- of $\{H_{1,k}\}_{k = 1}^K$ are false. We use this assumption about $\{H_{1,k}\}_{k = 1}^K$ for design purposes to ensure the Bayesian  FDR and average power calculations are non-trivial in all simulation repetitions; there are no restrictions on how many hypotheses are false when using posterior probabilities to assess a collection of A/B tests.  We note that one could also use our design framework with a model $\Psi$ such that all $\{H_{1,k}\}_{k = 1}^K$ are true or false in \emph{some} simulation repetitions $r = 1, \dots, m$. 

\begin{algorithm}
\caption{Discovery Classification with Sampling Distributions}
\label{alg.int}

\begin{algorithmic}[1]
\setstretch{1.1}
\Procedure{Classify}{$f(\cdot)$, $g(\cdot)$, $\boldsymbol{\delta}_L$, $\boldsymbol{\delta}_U$, $n$, $c$, $\boldsymbol{\gamma}$, $K$, $p(\boldsymbol{\eta})$, $m$, $\Psi$}
\For{$r$ in 1:$m$}
    \State Generate $\boldsymbol{\eta}_{r} \sim \Psi$ 
    \State Generate $\boldsymbol{y}^{_{(n)}}_{r} \sim f(\boldsymbol{y}; \boldsymbol{\eta}_{A,r}), f(\boldsymbol{y}; \boldsymbol{\eta}_{B,r})$ 
    \State $v_r \leftarrow 0$; $s_r \leftarrow 0$; $t_r \leftarrow 0$
    \For{$k$ in 1:$K$}
    \State Compute estimate $\widehat{Pr}(H_{1,k} ~| ~\boldsymbol{y}^{_{(n)}}_{r})$ 
    \If {$g_k(\boldsymbol{\eta}_{r}) \notin (\delta_{L,k}, \delta_{U, k})$}
    \If {$\widehat{Pr}(H_{1,k} ~| ~\boldsymbol{y}^{_{(n)}}_{r}) \ge \gamma_k$}
    \State $v_r \leftarrow v_r + 1$
    \EndIf
    \Else 
    \State $t_r \leftarrow t_r + 1$
    \If {$\widehat{Pr}(H_{1,k} ~| ~\boldsymbol{y}^{_{(n)}}_{r}) \ge \gamma_k$}
    \State $s_r \leftarrow s_r + 1$
    \EndIf
    \EndIf
    \EndFor
    \EndFor
    \State \Return $\{v_r\}_{r = 1}^m$, $\{s_r\}_{r = 1}^m$, and $\{t_r\}_{r = 1}^m$
\EndProcedure

\end{algorithmic}
\end{algorithm}

Line 4 of Algorithm \ref{alg.int} generates a sample $\boldsymbol{y}^{_{(n)}}_{r}$ in each simulation repetition given the value for $\boldsymbol{\eta}_{r}$ drawn from $\Psi$ in Line 3. The collection of $\{\{\widehat{Pr}(H_{1, k}~| ~\boldsymbol{y}^{_{(n)}}_{r})\}_{k=1}^K\}_{r = 1}^m$ values from Line 7 across all hypotheses and simulation repetitions estimates the joint sampling distribution of posterior probabilities under the model $\Psi$. The operating characteristics of the Bayesian  FDR defined in (\ref{eq:fdr.1})  and average power defined in (\ref{eq:avg.pwr})  can respectively be estimated as
\begin{equation}\label{eq:oc.est}
\dfrac{1}{m}\sum_{r=1}^m\dfrac{v_r}{\max\{v_r + s_r, 1\}}  ~~~ \text{and} ~~~ \dfrac{1}{m}\sum_{r=1}^m\dfrac{s_r}{\max\{t_r, 1\}},
\end{equation} 
where $v_r$, $s_r$, and $t_r$ are the number of false discoveries, true discoveries, and true hypotheses for simulation repetition $r$ defined in Algorithm \ref{alg.int}.
The denominators in (\ref{eq:oc.est}) avoid division by 0. Based on Algorithm \ref{alg.int}, an $(n, \boldsymbol{\gamma})$ combination is suitable if the Bayesian  FDR estimate in (\ref{eq:oc.est}) is at most $q$ and the average power estimate in (\ref{eq:oc.est}) is at least $1 - \beta$.

We now separately consider how the marginal sampling distribution of $Pr(H_{1, k}~|~ \boldsymbol{y}^{_{(n)}})$ for each hypothesis $H_{1, k}$ behaves with respect to $n$. The marginal sampling distribution of $Pr(H_{1, k}~|~ \boldsymbol{y}^{_{(n)}})$ has at least one of the following two components. The first component is based on $Pr(H_{1, k}~|~ \boldsymbol{y}^{_{(n)}}_r)$ for $\boldsymbol{\eta}_{r} \sim \Psi$ such that $H_{1, k}$ is true. As $n \rightarrow \infty$, this component of the marginal sampling distribution generally converges to a point mass at 1 \citep{vaart1998bvm}. The second component is based on $Pr(H_{1, k}~|~ \boldsymbol{y}^{_{(n)}}_r)$ for $\boldsymbol{\eta}_{r} \sim \Psi$ such that $H_{1, k}$ is false. When all those $\boldsymbol{\eta}_{r}$ values are such that $\theta_{r,k} = g_k(\boldsymbol{\eta}_{r})$ equals $\delta_{L,k}$ or $\delta_{U, k}$, this component of the marginal sampling distribution for $H_{1,k}$ converges to the standard uniform distribution as $n \rightarrow \infty$ under weak conditions \citep{bernardo2009bayesian}.

The optimal decision thresholds $\boldsymbol{\gamma}$ that bound the estimated Bayesian  FDR in (\ref{eq:oc.est}) by $q$ can be found using a defined optimization scheme. We illustrate the use of several constrained  optimization  schemes that select $\boldsymbol{\gamma}$ to maximize average power in Section \ref{sec:ex}. As $n$ increases, so do the posterior probabilities in the components of the marginal sampling distributions where $H_{1,k}$ is true, and it becomes easier to distinguish between the true and false hypotheses. The optimal $\boldsymbol{\gamma}$ values for a given optimization scheme typically decrease alongside $n$; however, this decreasing trend is attenuated as $n \rightarrow \infty$ since the components of the marginal sampling distributions of posterior probabilities where $H_{1,k}$ is true converge to 1. Even as the optimal $\boldsymbol{\gamma}$ values change, the estimated average power in (\ref{eq:oc.est}) increases to 1 as $n$ increases. The optimal choices for $n$ and $\boldsymbol{\gamma}$ are therefore intrinsically linked when designing A/B tests based on the Bayesian  FDR and average power. 

These choices for $n$ and $\boldsymbol{\gamma}$ are informed by exploring the joint sampling distribution of posterior probabilities under $\Psi$ to estimate the operating characteristics in (\ref{eq:oc.est}). While we can examine multiple $\boldsymbol{\gamma}$ values using the same estimate of the sampling distribution for a given sample size $n$, the procedure in Algorithm \ref{alg.int} requires independent implementation for each $n$ value we consider. This process is often computationally intensive, but we could reduce the computational burden by repurposing the results from Algorithm \ref{alg.int} for previously considered sample sizes to estimate operating characteristics for new $n$ values. This process would allow us to explore new $(n, \boldsymbol{\gamma})$ combinations without conducting additional simulations. We begin the development of such a method for the design of A/B tests in Section \ref{sec:proxy}. 


\section{A Proxy to the Joint Sampling Distribution}\label{sec:proxy}

To motivate our design methods proposed in Section \ref{sec:power}, we develop a proxy to the joint sampling distribution of posterior probabilities. These proxies are required for the theory that substantiates our methodology, but our methods do not directly use them. Instead, we estimate the true joint sampling distribution of posterior probabilities by generating data $\boldsymbol{y}^{_{(n)}}$ using the straightforward process in Algorithm \ref{alg.int}. Our proxies leverage the regularity conditions listed in Appendix A of the supplement. Appendix A.1 details the four necessary assumptions to invoke the Bernstein-von Mises (BvM) theorem \citep{vaart1998bvm}.  The first three assumptions are also required for the asymptotic normality of the maximum likelihood estimator (MLE) \citep{vaart1998bvm}, and the MLE regularity conditions are listed in Appendix A.2. 

By the BvM theorem, a large-sample approximation to the posterior of $\boldsymbol{\theta}~| ~\boldsymbol{y}^{_{(n)}}_{r}$ is
  \begin{equation}\label{eq:bvm}
  \mathcal{N}\left(\hat{\boldsymbol{\theta}}^{_{(n)}}_r, \mathcal{I}(\boldsymbol{\theta}_r)^{-1}/n\right),
\end{equation}
where $\hat{\boldsymbol{\theta}}^{_{(n)}}_{r}$ is the maximum likelihood estimate of $\boldsymbol{\theta}$, $\mathcal{I}(\cdot)$ is the Fisher information with respect to the targets of inference, and $\boldsymbol{\theta}_{r} = \{\theta_{r,k}\}_{k=1}^K = \{g_k(\boldsymbol{\eta}_{r})\}_{k=1}^K$. We approximate the joint posterior of $\boldsymbol{\theta}$ in (\ref{eq:bvm}) to allow for an arbitrary dependence structure between its components.  The approximate sampling distribution of the MLE $\hat{\boldsymbol{\theta}}^{_{(n)}} ~|~ \boldsymbol{\eta} = \boldsymbol{\eta}_{r}$ is $\mathcal{N}(\boldsymbol{\theta}_r, n^{-1}\mathcal{I}(\boldsymbol{\theta}_r)^{-1})$ under the regularity conditions in \citet{vaart1998bvm}. To develop our proxy used for theoretical purposes, a single realization from this $K$-dimensional multivariate normal distribution could be generated using conditional cumulative distribution function (CDF) inversion and a point $\boldsymbol{u} = \{u_k \}_{k=1}^K \in [0,1]^K$. We could obtain the first component $\hat{\theta}^{_{(n)}}_{r, 1}$ as the $u_{1}$-quantile of the sampling distribution of $\hat{\theta}^{_{(n)}}_{1} ~|~ \boldsymbol{\eta}_{r}$. For the remaining components, we could iteratively generate $\hat{\theta}^{_{(n)}}_{r, k}$ as the $u_{k}$-quantile of the sampling distribution of $\hat{\theta}^{_{(n)}}_{k} ~|~ \{\hat{\theta}^{_{(n)}}_{s} = \hat{\theta}^{_{(n)}}_{r, s}\}_{s=1}^{k-1}, \hspace*{0.1pt} \boldsymbol{\eta}_{r}$. 


Implementing this process with a sequence of $m$ points $\{\boldsymbol{u}_{r}\}_{r = 1}^m \sim \mathcal{U}\left([0,1]^{K}\right)$ simulates a sample from the approximate sampling distribution of $\hat{\boldsymbol{\theta}}^{_{(n)}}$ according to $\Psi$. We could substitute this sample $\{ \hat{\boldsymbol{\theta}}^{_{(n)}}_{ r}\}_{r=1}^m$ into the posterior approximation in (\ref{eq:bvm}) to yield a proxy sample of posterior probabilities. For a given simulation repetition $r$, the following probability is a large-sample proxy to the posterior probability that $H_{1,k}$ is true:
      \begin{equation}\label{eq:proxy}
p^{_{(n)}}_{r, k} = 
   \Phi\left(\dfrac{\delta_{U, k} - \hat{\theta}^{_{(n)}}_{r, k}}{\sqrt{n^{-1}\mathcal{I}_{k,k}(\boldsymbol{\theta}_r)^{-1}}}\right) - \Phi\left(\dfrac{\delta_{L, k} - \hat{\theta}^{_{(n)}}_{r, k}}{\sqrt{n^{-1}\mathcal{I}_{k,k}(\boldsymbol{\theta}_r)^{-1}}}\right),
\end{equation} 
where $\Phi(\cdot)$ is the standard normal CDF. The collection of $\{p^{_{(n)}}_{r, k}\}_{k = 1}^K$ values corresponding to $\{\boldsymbol{u}_{r}\}_{r = 1}^m \sim \mathcal{U}\left([0,1]^{K}\right)$ and $\{\boldsymbol{\eta}_{r}\}_{r = 1}^m \sim \Psi$ define our proxy to the joint sampling distribution of posterior probabilities.  Theory from \citet{hagar2024fast} can be extended to show that the total variation distance between the proxy sampling distribution and the true sampling distribution converges in probability to 0 as $n \rightarrow \infty$. However, the proxy and true sampling distributions could differ materially for finite $n$. 

Thus, the proxy sampling distribution only motivates our theoretical result in Theorem \ref{thm1}. This result guarantees that the logit of $p^{_{(n)}}_{r, k}~|~\{p^{_{(n)}}_{r, h}\}_{h = 1}^{k-1}$ is an approximately linear function of $n$ for each metric $k$.  For $k = 1$, the conditioning set $\{p^{_{(n)}}_{r, h}\}_{h = 1}^{k-1}$ is empty.  We later adapt this result to assess the operating characteristics of A/B tests across a wide range of sample sizes by estimating the true joint sampling distribution of posterior probabilities under $\Psi$ at only \emph{two} values of $n$. Each $p^{_{(n)}}_{r, k}$ value depends on the value for $\hat{\boldsymbol{\theta}}^{_{(n)}}_{r}$, which in turn depends on the sample size $n$, the parameter value $\boldsymbol{\eta}_{r}$, and the point $\boldsymbol{u}_{r}$. In Theorem \ref{thm1}, we fix both $\boldsymbol{\eta}_{r}$ and $\boldsymbol{u}_{r}$ to examine the behaviour of $p^{_{(n)}}_{r, k}~|~\{p^{_{(n)}}_{r, h}\}_{h = 1}^{k-1}$ as a deterministic function of $n$.

\begin{theorem}\label{thm1}
    For any $\boldsymbol{\eta}_{r} \sim \Psi$, let the model $f(\boldsymbol{y};\boldsymbol{\eta}_{r})$ satisfy the conditions in Appendix A.2 and the prior $p(\boldsymbol{\eta})$ satisfy the conditions in Appendix A.1.
    Define $\emph{logit}(x) = \emph{log}(x) - \emph{log}(1-x)$. We consider a given point $\boldsymbol{u}_{r} = \{u_{r,k}\}_{k=1}^K \in [0,1]^{K}$. For $k = 1, \dots, K$, the functions $p^{_{(n)}}_{r, k}$ in (\ref{eq:proxy}) are such that
    \vspace*{-5pt}
 \begin{enumerate}
     \item[(a)] $p^{_{(n)}}_{r, k}~|~\{p^{_{(n)}}_{r, h}\}_{h = 1}^{k-1} = \Phi\left(
           a_k(\delta_{U, k}, \boldsymbol{\theta}_{r})\sqrt{n} + b_k(\{u_{r,h}\}_{h=1}^k) 
         \right) - \Phi\left(
           a_k(\delta_{L, k}, \boldsymbol{\theta}_{r})\sqrt{n} + b_k(\{u_{r,h}\}_{h=1}^k) 
         \right)$, where $a_k(\delta_{k}, \boldsymbol{\theta}_{r}) = (\delta_{k} - \theta_{r,k})/\sqrt{\mathcal{I}_{k,k}(\boldsymbol{\theta}_r)^{-1}}$ and  $b_k(\cdot)$ is a not a function of $n$.
     \item[(b)] $\lim\limits_{n \rightarrow \infty} \dfrac{d}{dn}~\emph{logit}\left(p^{_{(n)}}_{r, k}~|~\{p^{_{(n)}}_{r, h}\}_{h = 1}^{k-1} \right)= (0.5 - \mathbb{I}\{\theta_{r,k} \notin (\delta_{L, k}, \delta_{U, k})\})\times\emph{min}\{a_k(\delta_{U, k}, \boldsymbol{\theta}_{r})^2, a_k(\delta_{L,k}, \boldsymbol{\theta}_{r})^2\} $. 
 \end{enumerate}
\end{theorem} 

We prove parts $(a)$ and $(b)$ of Theorem \ref{thm1} in Appendix B of the supplement. \citet{hagar2025economical} developed simplified theory for $p^{_{(n)}}_{r, k}$ corresponding to a single hypothesis.  However, that result does not account for the dependence structure across multiple hypotheses and cannot be used to model \emph{joint} sampling distributions of posterior probabilities, hence the need for the extended theory.  We now consider the practical implications of Theorem \ref{thm1}. The limiting derivative in part $(b)$ is a constant that does not depend on $n$.  Moreover, these limiting derivatives do not depend on the point $\boldsymbol{u}_r$, which controls the dependence in the joint proxy sampling distribution. The limiting derivatives for $\text{logit}\left(p^{_{(n)}}_{r, k}\right)$ and $\text{logit}\left(p^{_{(n)}}_{r, k}~|~\{p^{_{(n)}}_{r, h}\}_{h = 1}^{k-1}\right)$ are therefore the same.  For the probabilities corresponding to each metric in the joint sampling distribution, the linear approximation to $l^{_{(n)}}_{r, k} = \text{logit}(p^{_{(n)}}_{r, k}~|~\{p^{_{(n)}}_{r, h}\}_{h = 1}^{k-1})$ as a function of $n$ is a good global approximation for sufficiently large sample sizes. Moreover, this linear approximation should be locally suitable for a range of sample sizes. 

It follows that the  (conditional) quantiles of the sampling distribution of $l^{_{(n)}}_{r, k}$ change linearly as a function of $n$ when $\boldsymbol{\theta}_{r}$ is held constant across simulation repetitions. In Section \ref{sec:power}, we exploit and adapt this linear trend in the proxy sampling distribution to flexibly model the logits of posterior probabilities as linear functions of $n$ when independently simulating samples $\boldsymbol{y}^{_{(n)}}$ according to  $\boldsymbol{\theta}_{r} \sim \Psi$ as in Algorithm \ref{alg.int}. To ensure our method performs well with finite sample sizes, we only use the limiting slopes from Theorem \ref{thm1} to initialize our approach. The good performance of our simulation-based method, which empirically estimates linear functions from the data, is illustrated in Section \ref{sec:ex}.

   \section{Economical Assessment of the FDR and Power}\label{sec:power}

   We generalize the results from Theorem \ref{thm1} to develop the procedure in
   Algorithm \ref{alg3}. This procedure allows practitioners to efficiently explore the ($n, \boldsymbol{\gamma}$)-space to find the $(n, \boldsymbol{\gamma})$ combination under a given optimization scheme that minimizes the sample size while satisfying criteria for the Bayesian  FDR and average power. Algorithm \ref{alg3} is economical because we estimate the true joint sampling distribution of posterior probabilities via simulation at only two sample sizes: $n_0$ and $n_1$. The initial sample size $n_0$ is an input for Algorithm \ref{alg3}, and it can be selected based on the anticipated budget or timeline for the A/B test. We note that Algorithm \ref{alg3} details a general application of our methodology, and we later describe potential modifications.

\begin{algorithm}
\caption{Procedure to Determine Optimal Sample Size and Decision Thresholds}
\label{alg3}

\begin{algorithmic}[1]
\setstretch{1.25}
\Procedure{Optimize}{$f(\cdot)$, $g(\cdot)$, $\boldsymbol{\delta}_L$, $\boldsymbol{\delta}_U$, $p(\boldsymbol{\eta})$, $c$, $q$, $\beta$, $m$, $\Psi$, $n_0$}
\State Compute $\{\{\widehat{Pr}(H_{1,k} ~| ~\boldsymbol{y}^{_{(n_0)}}_{r})\}_{k=1}^K\}_{r=1}^m$ via Algorithm \ref{alg.int} and their logits  $\{\{\hat{l}^{_{(n_0)}}_{r, k}\}_{k=1}^K\}_{r=1}^m$
\For{$r$ in $1$:$m$}
\For{$k$ in $1$:$K$}
\State Use the line passing through $(n_0, \hat{l}^{_{(n_0)}}_{r, k})$ with the slope from Theorem \ref{thm1} to get $\hat{l}^{_{(n)}}_{r, k}$ for other $n$
\EndFor
\EndFor
\State Find $n_1$, the smallest $n$ such that Algorithm \ref{alg.int} with the expits of $\{\{\hat{l}^{_{(n)}}_{r, k}\}_{k=1}^K\}_{r=1}^m$ yields  average power \linebreak \hspace*{12pt} $\ge 1 - \beta$, where $\boldsymbol{\gamma}_n$ are the optimal thresholds  such that $\widehat{\text{FDR}}$ $\le q$
\State Compute $\{\{\widehat{Pr}(H_{1,k} ~| ~\boldsymbol{y}^{_{(n_1)}}_{r})\}_{k=1}^K\}_{r=1}^m$ via Algorithm \ref{alg.int} and their logits  $\{\{\hat{l}^{_{(n_1)}}_{r, k}\}_{k=1}^K\}_{r=1}^m$
\For{$d$ in $1$:$m$}
\For{$k$ in $1$:$K$}
\State Let $\hat{l}^{_{(n_0)}}_{d, k}$ and $\hat{l}^{_{(n_1)}}_{d, k}$ be the $d^{\text{th}}$ order statistics of $\{\hat{l}^{_{(n_0)}}_{r, k} \}_{r=1}^m$ and $\{\hat{l}^{_{(n_1)}}_{r, k} \}_{r=1}^m$
\State Let $r$ be the index of the sample $\boldsymbol{y}^{_{(n_1)}}_{r}$ corresponding to $\hat{l}^{_{(n_1)}}_{d, k}$
\State Now use the line $\hat{L}^{_{(n)}}_{r, k}$ passing through $(n_0, \hat{l}^{_{(n_0)}}_{d, k})$ and $(n_1, \hat{l}^{_{(n_1)}}_{d, k})$ to get  $\hat{l}^{_{(n)}}_{r, k}$ for other $n$
\EndFor
\EndFor
\State Find $n_2$, the smallest $n$ such that Algorithm \ref{alg.int} with the expits of $\{\{\hat{l}^{_{(n)}}_{r, k}\}_{k=1}^K\}_{r=1}^m$  yields  average  power \linebreak \hspace*{12pt} $\ge 1 - \beta$, where $\boldsymbol{\gamma}_n$ are the optimal thresholds  such that  $\widehat{\text{FDR}}$  $\le q$
 \State \Return $n_2$ as recommended $n$ and $\boldsymbol{\gamma}_{n_2}$ as $\boldsymbol{\gamma}$

\EndProcedure

\end{algorithmic}
\end{algorithm}

We now elaborate on several steps of Algorithm \ref{alg3}. In Line 2, we estimate posterior probabilities at the sample size $n_0$ by simulating data as in Algorithm \ref{alg.int}. These posterior probabilities can flexibly be computed using any computational or analytical approximation method. If a computational method is used to generate posterior samples, we recommend calculating posterior probabilities using a nonparametric kernel density estimate of the posterior so that the logits of all probabilities are finite. The notation $\hat{l}^{_{(n)}}_{r, k}$ is also introduced in Line 2. These estimated logits from the true joint sampling distribution leverage independently generated samples $\boldsymbol{y}^{_{(n)}}_{r}$ from $\Psi$ for each simulation repetition $r$. Because we estimate the joint sampling distribution of posterior probabilities, the resulting estimated logits allow us to model logits of the \emph{conditional} posterior probabilities considered in Section \ref{sec:proxy}. 

Unlike for $l^{_{(n)}}_{r, k}$ from the proxy sampling distribution in Theorem \ref{thm1}, there is no relationship between the $\hat{l}^{_{(n)}}_{r, k}$ values corresponding to two different sample sizes that happen to have the same indices for $r$ and $k$. In Line 5 of Algorithm \ref{alg3}, we construct linear approximations to the logits of posterior probabilities as a function of $n$ using the limiting slopes from Theorem \ref{thm1}. For moderate $n$, the limiting slopes for $l^{_{(n)}}_{r, k}$ may not be accurate since Theorem \ref{thm1} relies on asymptotic theory. Thus, we only use those slopes to initialize our method.

We choose the next sample size $n_1$ at which to estimate posterior probabilities in Line 8 of Algorithm \ref{alg3}. Recall that the optimal decision thresholds $\boldsymbol{\gamma}$ are linked with $n$ as described in Section \ref{sec:prelim}. To account for this phenomenon, we run an optimization scheme to find the optimal $\boldsymbol{\gamma}$ for which $\widehat{\text{FDR}}$, the estimate of the Bayesian  FDR defined in (\ref{eq:oc.est}), is at most $q$. This optimization scheme can be as simple or as complex as the practitioner desires; we elaborate on several potential schemes in Section \ref{sec:ex}. Line 8 introduces a subscript $n$ for the decision thresholds to emphasize this dependence on the sample size. Given $\boldsymbol{\gamma}_n$ for a particular $n$, we can determine whether this sample size achieves an average power of at least $1- \beta$. The sample size $n_1$ is the smallest $n$ such that the average power is sufficiently large. We recommend finding this sample size using binary search. We emphasize that a modified version of Algorithm \ref{alg.int} based on the linear approximations from Line 5 of Algorithm \ref{alg3} can be efficiently implemented for new sample sizes without computing additional posterior probabilities. 


In Lines 10 to 16 of Algorithm \ref{alg3}, we construct linear approximations to logits of posterior probabilities that are less reliant on large-sample results. These approximations are obtained separately for each metric $k$ by fitting a line through the same order statistic of the marginal sampling distribution at the sample sizes $n_0$ and $n_1$. Because Theorem \ref{thm1} ensures that the limiting slopes of $\text{logit}\left(p^{_{(n)}}_{r, k}\right)$ and $\text{logit}\left(p^{_{(n)}}_{r, k}~|~\{p^{_{(n)}}_{r, h}\}_{h = 1}^{k-1}\right)$ from the proxy sampling distribution are the same, we use the \emph{marginal} sampling distributions to estimate slopes for the \emph{conditional} logits. To maintain the proper level of dependence in the joint sampling distribution of posterior probabilities, we group the lines $\hat{L}^{_{(n)}}_{r, k}$ from Line 14 across metrics based on the sample $\boldsymbol{y}^{_{(n_1)}}_{r}$ that defines the linear approximations. Given the linear trend in the proxy sampling distribution quantiles discussed in Section \ref{sec:proxy}, it is reasonable to construct these linear approximations based on order statistics of estimates of the true sampling distributions when the $\boldsymbol{\theta}_{r}$ value is similar for all $\boldsymbol{\eta}_{r} \sim \Psi$. 

However, we recommend defining $\Psi$ as a mixture of various submodels, each of which gives rise to a different $\boldsymbol{\theta}_{r}$ value, to accommodate uncertainty about which hypotheses are false. This submodel approach is demonstrated in Section \ref{sec:ex}. In that case, the process in Lines 10 to 16 can be modified.  We instead split the logits of the posterior probabilities for each $n$ value into subgroups based on their submodels in $\Psi$ before constructing the linear approximations. We use this set of linear approximations to obtain the optimal sample size $n_2$ and its associated decision thresholds in Line 17. 

We provide a theoretical guarantee for the performance of Algorithm \ref{alg3} below. Note the recommended sample size $n$ is a function of the bound $q$ for the Bayesian FDR, the target value $1 - \beta$ for average power, the probability model $\Psi$, and the prior $p(\boldsymbol{\eta})$. To allow the recommended $n$ value to approach $\infty$ in our theoretical results, we introduce a sequence for average power $(1 - \beta_n, \dots)$ that increases to 1.  Corollary \ref{cor1} details how the Bayesian FDR and average power for the design defined using the $(n, \boldsymbol{\gamma})$ combination returned by Algorithm \ref{alg3} with inputs $q$ and $1 - \beta_n$ respectively converge to $q$ and $1 - \beta_{\color{purple}n}$ as the recommended value of $n \rightarrow \infty$.

\begin{corollary}\label{cor1}
    Let the conditions for Theorem \ref{thm1} be satisfied. Let the Bayesian FDR and average power be defined as in (\ref{eq:fdr.1}) and (\ref{eq:avg.pwr}), respectively. Consider a design with the $(n, \boldsymbol{\gamma})$ combination recommended by Algorithm \ref{alg3} with inputs $q$ and $1 - \beta_n$. As the recommended $n \rightarrow \infty$, this design is such that 
    $$\left\lvert\mathbb{E}_{\boldsymbol{\eta} \sim \Psi}\left\{\mathbb{E}\left[\frac{V}{V+S}\right]\right\} - q \right\rvert \xrightarrow{P} 0 ~~\text{and}~~\left\lvert\mathbb{E}_{\boldsymbol{\eta} \sim \Psi}\left\{\mathbb{E}\left[\frac{S}{T}\right]\right\} - (1 - \beta_{\color{purple}n}) \right\rvert \xrightarrow{P} 0.$$
\end{corollary}

We prove Corollary \ref{cor1} in Appendix C of the online supplement. While Algorithm \ref{alg3} leverages asymptotic results, its linear approximations are constructed by simulating sampling distributions of posterior probabilities at finite sample sizes $n$. With finite sample sizes that are sufficiently large, Algorithm \ref{alg3} with fixed inputs $q$ and $1 - \beta$ thus recommends $(n, \boldsymbol{\gamma})$ combinations such that the Bayesian FDR  approximates $q$ and average power approximates $1 - \beta$.  In Section \ref{sec:ex}, we consider the performance of this algorithm for an example based on a real A/B test involving 5 metrics.

\section{Optimizely Example}\label{sec:ex}

Optimizely is a company that provides software to help other businesses conduct online controlled experiments. Optimizely also advocates for improving their own online experiences; \citet{siroker2013ab} described an A/B test that Optimizely ran amidst a 2012 redesign of their website. In this section, we illustrate the use of our methods to design an experiment with $K = 5$ of the metrics considered by Optimizely during their redesign. The number of metrics we consider ($K = 5$) is typical for A/B tests \citep{ghosh2021experimental} and aligns with the rough rule of thumb provided by \citet{kohavi2020trustworthy}. Each metric corresponds to a distinct binary outcome for visitor $i = 1,\dots, n$:
\begin{itemize}[itemsep = -5pt, topsep = 0pt]
    \item $y_{1i}$: \emph{engaged} with the experiment by clicking anywhere on the homepage.
    \item $y_{2i}$: used the Optimizely \emph{editor} tool.
    \item $y_{3i}$: visited Optimizely's \emph{pricing} page.
    \item $y_{4i}$: triggered the \emph{dialog} to create an Optimizely account.
    \item $y_{5i}$: successfully \emph{created} an Optimizely account.
\end{itemize}

\begin{figure}[!tb]
		\includegraphics[width = \textwidth]{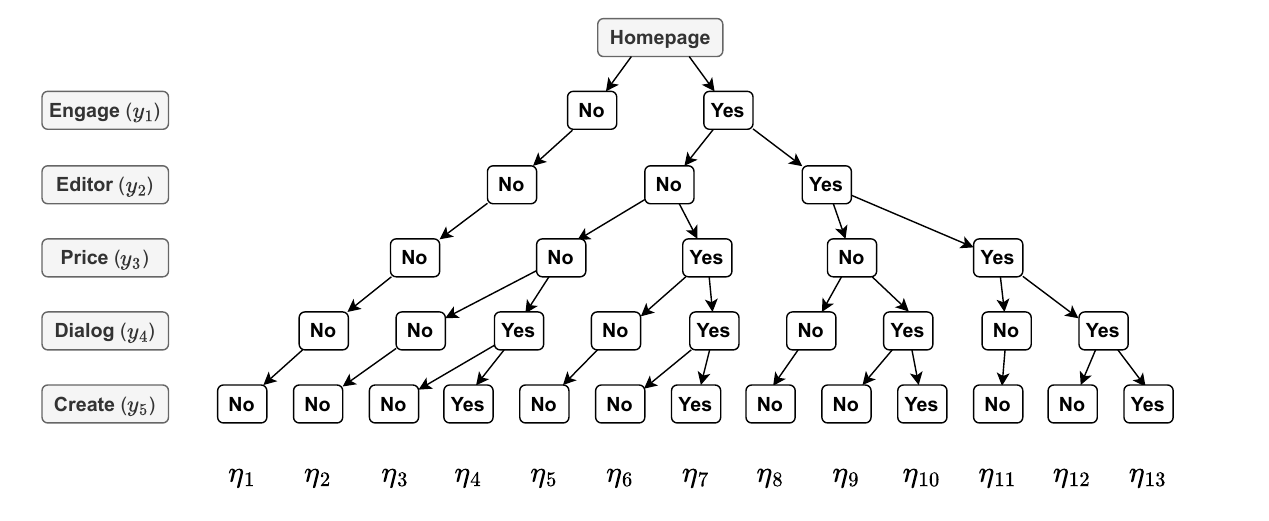} 

		\caption{\label{fig:opt} Visualization of the multinomial model that characterizes the admissible combinations of the five binary outcomes from the Optimizely experiment. The multinomial categories are indexed by the probabilities $\{\eta_v\}_{v=1}^{13}$. } 
	\end{figure}

     This example is interesting because these five binary outcomes are not independent. In particular, a visitor cannot experience the final four outcomes unless they engage with the experiment. Moreover, the account creation dialog related to the fourth outcome must be shown for a visitor to successfully create an Optimizely account. We circumvent the need to directly specify the dependence structure between the Optimizely outcomes by jointly modeling all five binary variables using a multinomial model.  Figure \ref{fig:opt} visualizes this multinomial model, which has 13 probabilities associated with the five outcomes. We first introduce the model for a single treatment group and later introduce subscripts for the group $j$, where group A is the original site and group B is the redesign. In a particular group, the marginal probability of a visitor attaining the $k^{\text{th}}$ binary outcome is represented by $\pi_k, ~k = 1, \dots, 5$. These marginal probabilities can be expressed as sums of various multinomial probabilities $\{\eta_{v}\}_{v=1}^{13}$ from Figure \ref{fig:opt}:  $\pi_1 = 1 - \eta_1$, $\pi_2 = \sum_{v=8}^{13}\eta_v$, $\pi_3 = \sum_{v=5}^7\eta_v + \sum_{v=11}^{13}\eta_v$, $\pi_4 = \sum_{v=1}^4\eta_{3v} + \sum_{v=1}^4\eta_{3v+ 1}$, $\pi_5 = \sum_{v=1}^4\eta_{3v+ 1}$.

      

For the $k^{\text{th}}$ binary outcome, \citet{siroker2013ab} considered the relative difference (i.e., the \emph{lift}) between the probabilities of a visitor attaining that outcome in group B vs.\ group A. 
We use lift as the metric of interest for this example:
\begin{equation*}\label{eq:lift}
    \theta_k = \dfrac{\pi_{B, k} - \pi_{A, k}}{\pi_{A, k}},
\end{equation*}
where the first subscript now denotes the group $j = A, B$. Our hypotheses of interest are $H_{0, k}: \theta_k \le 0$ vs. $H_{1,k}: \theta_k > 0$. That is, $(\delta_{L, k}, \delta_{U, k}) = (0, \infty)$ for all metrics $k = 1, \dots, 5$. Thus, this experiment aims to demonstrate that the redesigned website is superior to the original one with respect to the various binary outcomes. We emphasize that the hypotheses $\{H_{1,k}\}_{k = 1}^5$ are not independent due to the relationship between $\{\pi_{j, k}\}_{k = 1}^5$ and $\boldsymbol{\eta}_j = \{\eta_{j,v} \}_{v = 1}^{13}$. 

To implement our method, we must define the probability model $\Psi$ that characterizes data generation. We specify this model by choosing various sets of multinomial models. Each set of models defines separate multinomial probabilities for the two groups. Specifying various sets of multinomial models allows us to accommodate uncertainty in which hypotheses $H_{1,k}$ are false. It also allows us to account for uncertainty in effect sizes for a given set of true hypotheses, but we do not utilize that flexibility in this example. For all sets of multinomial models we consider, the marginal probabilities for group A align with the estimated marginal probabilities for the original site reported by \citet{siroker2013ab}: $\hat{\pi}_{A, 1} = 0.489, \hat{\pi}_{A, 2} = 0.230, \hat{\pi}_{A, 3} = 0.156, \hat{\pi}_{A, 4} = 0.047, \hat{\pi}_{A, 5} = 0.032$. Optimizely only provided the marginal probabilities, but we must specify the 13 multinomial probabilities $\{\eta_{A,v} \}_{v = 1}^{13}$ for group A. Below, we overview a method to choose these multinomial probabilities and those for group B. 

We choose each set of multinomial probabilities for both groups using linear programming \citep{guenin2014gentle}. Linear programming allows us to construct multinomial models that satisfy the following objectives. First, the probabilities $\sum_{v=1}^{13}\eta_{j, v}$ must sum to 1 in both groups. Second, we must attain specified values for the marginal probabilities $\{\pi_{A, k}\}_{k =1}^5$ in group A. Third, we need the marginal probabilities $\{\pi_{B, k}\}_{k =1}^5$ in group B to achieve targets for lift based on anticipated values for $\{\theta_k\}_{k = 1}^5$. These anticipated values determine which hypotheses in $\{H_{1,k}\}_{k =1}^5$ are true along with the effect sizes corresponding to the true hypotheses. We elaborate on this method to select a set of multinomial models in Appendix D.1 of the supplement. 


For a general A/B test, there are $ K \choose k$ ways to classify $k$ of the $K$ hypotheses as false. In this Optimizely experiment, there are $\sum_{k=1}^{4}{ 5 \choose k} = 30$ unique classifications of $\{H_{1,k}\}_{k =1}^5$ such that some but not all of the hypotheses are false. We choose to define $\Psi$ by specifying 30 sets of multinomial models in Appendix D.1. We index these submodels by introducing a subscript to the notation $\Psi_{\boldsymbol{\Lambda}}$, where $\varnothing \subsetneq \boldsymbol{\Lambda} \subsetneq \{1, 2, 3, 4, 5\}$ is the set of hypotheses that are false. For the submodel $\Psi_{\boldsymbol{\Lambda}}$, the multinomial models for each group ensure that $\theta_k = 0$ for $k \in \boldsymbol{\Lambda}$ and $\theta_k = 10\%$ otherwise. It follows that $H_{1,k}$ is false in this submodel if and only if $k \in \boldsymbol{\Lambda}$. For illustration, we define the model $\Psi$ input into Algorithm \ref{alg3} as an equally weighted mixture of these 30 submodels.

We now apply the method in Algorithm \ref{alg3} with balanced sample sizes ($c=1$) along with criteria for the Bayesian  FDR and average power defined by $q = 0.05$ and $\beta=0.2$. For now, we find optimal decision thresholds under the constraint that $\boldsymbol{\gamma} = \gamma \times \mathbf{1}_K$. This constraint gives rise to a simple optimization scheme to select $\gamma$: the optimal decision threshold is the smallest probability such that $\widehat{\text{FDR}} \le q$. We choose an initial sample size of $n_0 = 1.2\times10^4$ using information from Optimizely's 2012 experiment \citep{siroker2013ab}. We implemented our method with $10^3$ simulation repetitions for each of the 30 submodels, giving rise to a total of $m = 3\times10^4$ simulation repetitions. The prior $p(\boldsymbol{\eta})$ is comprised of independent, diffuse Dirichlet$(\mathbf{1}_{13})$ priors for the multinomial probabilities $\boldsymbol{\eta}_j$ in each group. These distributions are conjugate priors for $\boldsymbol{\eta}_A$ and $\boldsymbol{\eta}_B$, which we use to estimate the joint posterior of $\boldsymbol{\theta}$ via simulation. 

Algorithm \ref{alg3} with these inputs returned a recommended $(n_A, \gamma)$ combination of $(13328,$ $0.9411)$. All sample size recommendations in this section pertain to the sample size per group (i.e., $n_A = n_B$). It took roughly 3 minutes to obtain this recommendation on a parallel computing server with 72 cores. While each posterior takes less than one second to approximate, we must approximate many posteriors to reliably characterize the uncertainty regarding which hypotheses are false. Our method could be expedited by considering multivariate analytical approximations to the posterior of $\boldsymbol{\theta}$. It would have taken over 20 minutes using the same computing resources to explore the sample size space using binary search instead of Algorithm \ref{alg3}; this discrepancy in runtime scales logarithmically with the recommended sample size.

To confirm the strong performance of our method, we estimated the joint sampling distribution of posterior probabilities for $n_A = 13328$ under the model $\Psi$ with $m = 9.9 \times 10^4$ simulation repetitions. Using a decision threshold of $\gamma = 0.9411$, the Bayesian  FDR and average power in (\ref{eq:oc.est}) were estimated as 0.0497 and 0.7992. This good performance occurs under the assumption that $\boldsymbol{\eta} \sim \Psi$. If $\Psi$ is misspecified such that the effect sizes are generally overstated, the true Bayesian  FDR and average power are likely to exceed $q$ and not exceed $1 - \beta$, respectively. In contrast, a design informed by a model $\Psi$ that generally understates the effect sizes is likely to have a Bayesian  FDR and average power that outperform their respective criteria. We therefore advise that practitioners define $\Psi$ with respect to the minimum effect sizes they would like to detect.

To demonstrate the flexibility of our method with different models and the value of considering the dependence structure at the design stage, we implemented Algorithm \ref{alg3} with a modified version of $\Psi$ where the 5 five binary outcomes are generated independently according to the marginal probabilities from a given submodel $\Psi_{\boldsymbol{\Lambda}}$. The individual metrics in $\boldsymbol{\theta}$ were also analyzed using posteriors based on independent binomial models for each outcome and diffuse BETA(1,1) priors for each marginal probability. When ignoring the dependence structure between the 5 metrics, Algorithm \ref{alg3} returned a recommended $(n_A, \gamma)$ combination of $(14427, 0.9500)$ under the constraint that $\boldsymbol{\gamma} = \gamma \times \mathbf{1}_K$. Thus, when accounting for the positive dependence between these binary outcomes, we can use a less strict decision threshold, resulting in a materially smaller sample size recommendation. If the outcomes exhibit negative dependence, we generally need a larger decision threshold -- and sample size -- to satisfy criteria for both the Bayesian  FDR and average power. 


To illustrate how the number of false hypotheses impacts the sample size recommendation, we implement Algorithm \ref{alg3} where $\Psi$ is defined using only some of the 30 multinomial submodels. We again enforce the constraint that $\boldsymbol{\gamma} = \gamma \times \mathbf{1}_K$. We first consider the $(n_A, \gamma)$ recommendation where $\Psi$ is an equal mixture of the 5 submodels where only 1 hypothesis is false (i.e., where the cardinality $\lVert \boldsymbol{\Lambda} \rVert$ of the set $\boldsymbol{\Lambda}$ is 1). We separately obtain $(n_A, \gamma)$ recommendations for the 10 submodels where $\lVert \boldsymbol{\Lambda} \rVert = 2$, the 10 submodels where $\lVert \boldsymbol{\Lambda} \rVert = 3$, and the 5 submodels where $\lVert \boldsymbol{\Lambda} \rVert = 4$. Each recommendation was obtained via Algorithm \ref{alg3} with $m = 3\times10^4$, $q = 0.05$, and $\beta = 0.2$. In Table \ref{tab:fdr}, these results are compared to the $(n_A, \gamma)$ recommendation where $\Psi$ was defined by combining all 30 multinomial submodels in an equal mixture.

\begin{table}[tb]
\centering
\begin{tabular}{cccccc} 
\hline 
$\Psi$  & $n_A$   & $\boldsymbol{\gamma}$ \\ \hline
$\lVert \boldsymbol{\Lambda} \rVert = 1$  & 4627 & $0.7772 \times \mathbf{1}_5$                \\
$\lVert \boldsymbol{\Lambda}\rVert = 2$  & 9985  & $0.9053 \times \mathbf{1}_5$            \\
$\lVert \boldsymbol{\Lambda}\rVert = 3$ & 15035  & $0.9516 \times \mathbf{1}_5$                \\
$\lVert \boldsymbol{\Lambda}\rVert = 4$ & 19668 & $0.9737 \times \mathbf{1}_5$  \\       Combined  & 13328 & $0.9411 \times \mathbf{1}_5$   
\\       Combined  &  11106 &  $(0.9599, 0.9599, 0.9565, 0.9099, 0.9099)$ \\ \hline
\end{tabular}
\caption{Recommended sample sizes $n_A$ and decision thresholds $\gamma$ under probability models $\Psi$ that do (Combined) and do not ($\lVert \boldsymbol{\Lambda} \rVert = \{1, 2, 3, 4\}$) account for uncertainty in the number of false hypotheses.  Two optimization schemes are considered for the Combined $\Psi$.}
\label{tab:fdr}
\end{table}

Table \ref{tab:fdr} illustrates that the recommendations for the sample size and decision threshold generally increase along with the number of false hypotheses for given choices of $q$ and $\beta$. Thus, our initial application of Algorithm \ref{alg3} demonstrates the benefit of incorporating uncertainty with respect to which hypotheses are false. However, it may be infeasible to consider all $\sum_{k=1}^{K-1}{ K \choose k}$ ways to classify some of $\{H_{1,k}\}_{k =1}^K$ as false when $K$ is large. In that event, it would still be beneficial to consider a subset of these $\sum_{k=1}^{K-1}{ K \choose k}$ combinations of false hypotheses across different values of $k$.

Table \ref{tab:fdr} also demonstrates the value in considering more sophisticated optimization schemes to select $\boldsymbol{\gamma}$. For illustration, we also selected $\boldsymbol{\gamma}$ as the decision thresholds that ensure $\widehat{\text{FDR}} \le q$ while maximizing average power subject to the constraint $\lvert \gamma_k - \gamma_h \rvert \le 0.05$ for thresholds $k \ne h$. We detail our scheme to obtain the optimal $\boldsymbol{\gamma}_n$ for a given sample size $n$ in Appendix D.2. By optimizing the decision thresholds using more flexible constraints, we reduced the recommended sample size to $n_A = 11106$. It took about 5 minutes on a parallel computing server to obtain this optimal design; 2 of these minutes were spent using a single core to implement our optimization scheme to find $\boldsymbol{\gamma}_n$ at each $n$ explored. While many optimization schemes not considered in this example could also be used to select $\boldsymbol{\gamma}$, the computational burden of the optimization scheme should be considered prior to implementing Algorithm \ref{alg3}.

    Lastly, we compare our sample size recommendations to those prompted by a Bonferroni correction used to control the FWER of A/B tests in frequentist settings. To bound the FWER for 5 hypotheses by 0.05, the Bonferroni correction recommends using a significance level of 0.01 for each test. This significance level is similar to a decision threshold of $\gamma = 0.99$ in our context. A sample size of $n_A = 27956$ is recommended to ensure the average power across these 5 hypothesis tests is at least 80\% while bounding the FWER by 0.05. We emphasize that control of the (Bayesian)  FDR and FWER are not interchangeable. Yet by comparing this sample size recommendation to those in Table \ref{tab:fdr}, we illustrate that -- even with as few as $K = 5$ metrics -- practitioners can reduce the required sample size for their A/B tests if willing to use designs that bound the Bayesian  FDR instead of the FWER. The benefit of controlling the Bayesian FDR will be even more pronounced in cases where $K$ is very large. While it is rare for $K$ to be very large in online controlled experiments, our method could also be applied in contexts that typically involve a much larger number of comparisons.

\section{Discussion}\label{sec:disc}

In this paper, we put forward an economical framework to design Bayesian A/B tests involving multiple metrics while satisfying criteria for the Bayesian  FDR and average power. This framework recommends optimal sample sizes and decision thresholds that are tailored to the specified dependence structure between the hypothesis tests and accommodates uncertainty regarding which hypotheses are false. The efficiency of this framework stems from considering a proxy for the joint sampling distribution of posterior probabilities based on large-sample theory to motivate estimating the true sampling distributions at only two sample sizes. This approach significantly decreases the computational overhead required to design Bayesian A/B tests, making them much more attractive and accessible to practitioners that want to control the Bayesian  FDR.

The methods proposed in this paper could be extended in various aspects. Further work could develop more sophisticated and efficient optimization schemes to determine decision thresholds $\{\gamma_k\}_{k=1}^K$ that control the Bayesian  FDR. It may also be useful to consider design criteria for A/B tests other than the Bayesian  FDR or the FWER. This consideration could be framed using the three types of metrics defined by \citet{kohavi2020trustworthy} based on the evidence supporting metric improvement or deterioration in group B. In particular, positive and negative metrics respectively have statistically significant evidence of improvement and deterioration, and flat metrics have no statistically significant evidence of improvement or deterioration. \citet{kohavi2020trustworthy} recommended implementing a new online experience if all metrics are flat or positive, with at least one metric being positive. To classify the $k^{\text{th}}$ metric as positive, negative, or flat, we would need to compare $Pr(H_{1, k}~|~ data)$ to both lower and upper decision thresholds in an alternative design framework.

Moreover, multiple comparisons are conducted in sequential A/B tests \citep{bajpai2025extensible}. The FWER for sequential tests can be controlled using theory from group sequential designs \citep{jennison1999group}. However, this theory was developed with clinical applications in mind and does not account for the aims of certain sequential A/B tests. For example, businesses may want to compare two online experiences across different subgroups of users as data are collected. Standard methods for sequential design assume that data from previous stages are retained at future analyses. If the subgroup of interest changes over time, customized decision thresholds and sample sizes must be derived. Future research could consider the design of sequential A/B tests with subgroups.


 \section*{Supplementary Material}
These materials include a detailed description of the conditions for Theorem \ref{thm1}, the proofs of Theorem \ref{thm1} and Corollary \ref{cor1}, and additional context for the example in Section \ref{sec:ex}. The code to implement our methods and conduct our numerical studies is available online: \url{https://github.com/lmhagar/ABTestFDR}.

	\section*{Funding Acknowledgement}
	This work was supported by the Natural Sciences and Engineering Research Council of Canada (NSERC) by way of a PGS-D scholarship, a postdoctoral fellowship, and Grant RGPIN-2019-04212.
	



\bibliographystyle{chicago}


\begin{thebibliography}{}

\bibitem[\protect\citeauthoryear{Bajpai, Redmond, Fabijan, Arai, Tan, and Paul}{Bajpai et~al.}{2025}]{bajpai2025extensible}
Bajpai, V.~K., W.~Redmond, A.~Fabijan, B.~Arai, Y.~Tan, and P.~Paul (2025).
\newblock Extensible experimentation platform: Effective {A}/{B} test analysis at scale.
\newblock {\em Proceedings of the 22nd International Conference on Software Architecture\/}.

\bibitem[\protect\citeauthoryear{Benjamini and Hochberg}{Benjamini and Hochberg}{1995}]{benjamini1995controlling}
Benjamini, Y. and Y.~Hochberg (1995).
\newblock Controlling the false discovery rate: {A} practical and powerful approach to multiple testing.
\newblock {\em Journal of the Royal statistical society: series B (Methodological)\/}~{\em 57\/}(1), 289--300.

\bibitem[\protect\citeauthoryear{Benjamini and Yekutieli}{Benjamini and Yekutieli}{2001}]{benjamini2001control}
Benjamini, Y. and D.~Yekutieli (2001).
\newblock The control of the false discovery rate in multiple testing under dependency.
\newblock {\em Annals of Statistics\/}, 1165--1188.

\bibitem[\protect\citeauthoryear{Bernardo and Smith}{Bernardo and Smith}{2009}]{bernardo2009bayesian}
Bernardo, J.~M. and A.~F. Smith (2009).
\newblock {\em Bayesian Theory}, Volume 405.
\newblock John Wiley \& Sons.

\bibitem[\protect\citeauthoryear{Deng}{Deng}{2015}]{deng2015objective}
Deng, A. (2015).
\newblock Objective {B}ayesian two sample hypothesis testing for online controlled experiments.
\newblock In {\em Proceedings of the 24th International Conference on World Wide Web}, pp.\  923--928.

\bibitem[\protect\citeauthoryear{Deng, Hagar, Stevens, Xifara, and Gandhi}{Deng et~al.}{2024}]{deng2024metric}
Deng, A., L.~Hagar, N.~T. Stevens, T.~Xifara, and A.~Gandhi (2024).
\newblock Metric decomposition in a/b tests.
\newblock In {\em Proceedings of the 30th ACM SIGKDD Conference on Knowledge Discovery and Data Mining}, pp.\  4885--4895.

\bibitem[\protect\citeauthoryear{Deng, Xu, Kohavi, and Walker}{Deng et~al.}{2013}]{deng2013improving}
Deng, A., Y.~Xu, R.~Kohavi, and T.~Walker (2013).
\newblock Improving the sensitivity of online controlled experiments by utilizing pre-experiment data.
\newblock In {\em Proceedings of the 6th ACM International Conference on Web Search and Data Mining}, pp.\  123--132.

\bibitem[\protect\citeauthoryear{Ghosh}{Ghosh}{2021}]{ghosh2021experimental}
Ghosh, S. (2021).
\newblock {\em Experimental Approaches to Strategy and Innovation}.
\newblock Harvard University.

\bibitem[\protect\citeauthoryear{Guenin, K{\"o}nemann, and Tuncel}{Guenin et~al.}{2014}]{guenin2014gentle}
Guenin, B., J.~K{\"o}nemann, and L.~Tuncel (2014).
\newblock {\em A Gentle Introduction to Optimization}.
\newblock Cambridge University Press.

\bibitem[\protect\citeauthoryear{Hagar and Stevens}{Hagar and Stevens}{2024}]{hagar2024fast}
Hagar, L. and N.~T. Stevens (2024).
\newblock {Fast power curve approximation for posterior analyses}.
\newblock {\em Bayesian Analysis\/}, 1 -- 26 doi.org/10.1214/24--BA1469.

\bibitem[\protect\citeauthoryear{Hagar and Stevens}{Hagar and Stevens}{2025}]{hagar2025economical}
Hagar, L. and N.~T. Stevens (2025).
\newblock An economical approach to design posterior analyses.
\newblock {\em Journal of the American Statistical Association\/}, doi.org/10.1080/01621459.2025.2476221.

\bibitem[\protect\citeauthoryear{Hochberg and Tamhane}{Hochberg and Tamhane}{2009}]{hochberg2009multiple}
Hochberg, Y. and A.~C. Tamhane (2009).
\newblock {\em Multiple Comparison Procedures}.
\newblock Hoboken, NJ: John Wiley \& Sons.

\bibitem[\protect\citeauthoryear{Jennison and Turnbull}{Jennison and Turnbull}{1999}]{jennison1999group}
Jennison, C. and B.~W. Turnbull (1999).
\newblock {\em Group sequential methods with applications to clinical trials}.
\newblock CRC Press.

\bibitem[\protect\citeauthoryear{Kohavi, Tang, and Xu}{Kohavi et~al.}{2020}]{kohavi2020trustworthy}
Kohavi, R., D.~Tang, and Y.~Xu (2020).
\newblock {\em Trustworthy Online Controlled Experiments: A Practical Guide to A/B Testing}.
\newblock Cambridge University Press.

\bibitem[\protect\citeauthoryear{Larsen, Stallrich, Sengupta, Deng, Kohavi, and Stevens}{Larsen et~al.}{2024}]{larsen2024statistical}
Larsen, N., J.~Stallrich, S.~Sengupta, A.~Deng, R.~Kohavi, and N.~T. Stevens (2024).
\newblock Statistical challenges in online controlled experiments: A review of {A}/{B} testing methodology.
\newblock {\em The American Statistician\/}~{\em 78\/}(2), 135--149.

\bibitem[\protect\citeauthoryear{Luca and Bazerman}{Luca and Bazerman}{2021}]{luca2021power}
Luca, M. and M.~H. Bazerman (2021).
\newblock {\em The Power of Experiments: Decision Making in a Data-Driven World}.
\newblock MIT Press.

\bibitem[\protect\citeauthoryear{Roy}{Roy}{2001}]{roy2001design}
Roy, R.~K. (2001).
\newblock {\em Design of Experiments Using The Taguchi Approach: 16 Steps to Product and Process Improvement}.
\newblock John Wiley \& Sons.

\bibitem[\protect\citeauthoryear{Siroker and Koomen}{Siroker and Koomen}{2013}]{siroker2013ab}
Siroker, D. and P.~Koomen (2013).
\newblock {\em A/B Testing: The Most Powerful Way to Turn Clicks into Customers}.
\newblock John Wiley \& Sons.

\bibitem[\protect\citeauthoryear{Stevens and Hagar}{Stevens and Hagar}{2022}]{stevens2022comparative}
Stevens, N.~T. and L.~Hagar (2022).
\newblock Comparative probability metrics: Using posterior probabilities to account for practical equivalence in {A}/{B} tests.
\newblock {\em The American Statistician\/}~{\em 76\/}(3), 224--237.

\bibitem[\protect\citeauthoryear{Storey}{Storey}{2002}]{storey2002direct}
Storey, J.~D. (2002).
\newblock A direct approach to false discovery rates.
\newblock {\em Journal of the Royal Statistical Society Series B: Statistical Methodology\/}~{\em 64\/}(3), 479--498.

\bibitem[\protect\citeauthoryear{Thomke}{Thomke}{2020}]{thomke2020experimentation}
Thomke, S.~H. (2020).
\newblock {\em Experimentation Works: The Surprising Power of Business Experiments}.
\newblock Harvard Business Press.

\bibitem[\protect\citeauthoryear{van~der Vaart}{van~der Vaart}{1998}]{vaart1998bvm}
van~der Vaart, A.~W. (1998).
\newblock {\em Asymptotic Statistics}.
\newblock Cambridge Series in Statistical and Probabilistic Mathematics. Cambridge University Press.

\bibitem[\protect\citeauthoryear{Wang and Gelfand}{Wang and Gelfand}{2002}]{wang2002simulation}
Wang, F. and A.~E. Gelfand (2002).
\newblock A simulation-based approach to {B}ayesian sample size determination for performance under a given model and for separating models.
\newblock {\em Statistical Science\/}~{\em 17\/}(2), 193--208.

\bibitem[\protect\citeauthoryear{Yekutieli and Benjamini}{Yekutieli and Benjamini}{1999}]{yekutieli1999resampling}
Yekutieli, D. and Y.~Benjamini (1999).
\newblock Resampling-based false discovery rate controlling multiple test procedures for correlated test statistics.
\newblock {\em Journal of Statistical Planning and Inference\/}~{\em 82\/}(1-2), 171--196.

\end{thebibliography}

\end{document}